\documentclass[12pt]{article} 
\usepackage{amssymb,amsmath,fullpage}
\usepackage{url}
\usepackage{graphicx}

\newtheorem{lemma}{Lemma}

\def\calX{\mathcal{X}}
\def\calY{\mathcal{Y}}
\def\calC{\mathcal{C}}

\def\bbR{\mathbb{R}}
\def\bbN{\mathbb{N}}
\def\gmm{\mathrm{gmm}}
\def\calY{\mathcal{Y}}

\def\calC{\mathcal{C}}
\def\calP{\mathcal{P}}
\def\bbX{\mathbb{X}}
\def\bTheta{\mathbf{\Theta}}
\def\bOmega{\mathbf{\Omega}}
\def\calF{\mathcal{F}}
\def\st{{\ :\ }}

\def\AIC{\mathrm{AIC}}
\def\argmin{\mathrm{argmin}}

\def\floor#1{ {\lfloor #1 \rfloor}}
\def\ceil#1{ {\lceil #1 \rceil}}

\begin{document}

\author{Frank Nielsen\\
Sony Computer Science Laboratories, Inc.\\
3-14-13 Higashi Gotanda\\
141-0022 Shinagawa-ku, Tokyo, Japan\\
{\tt Frank.Nielsen@acm.org}\\
and
Richard Nock\\
NICTA\\
Sydney, Australia\\
{\tt Richard.Nock@nicta.com.au}
\thanks{This work was done while R. Nock was with Universit\'e des  
Antilles-Guyane - CEREGMIA, Campus de Schoelcher, 97233 Schoelcher,  
France.}
}

\title{Optimal  interval clustering: Application to Bregman clustering and statistical mixture learning}

 \maketitle

\begin{abstract}
We present a generic dynamic programming method to compute the optimal clustering of $n$ scalar elements into $k$ pairwise disjoint intervals. This case includes 1D Euclidean $k$-means, $k$-medoids, $k$-medians, $k$-centers, etc.
We extend the method to incorporate cluster size constraints and show how to choose the appropriate $k$ by model selection.
Finally, we illustrate and refine the method on two case studies: Bregman clustering and statistical mixture learning maximizing the complete likelihood.
\end{abstract}

\parindent 0cm {\bf Key words}:
Clustering, dynamic programming, $k$-means, Bregman divergences, statistical mixtures, exponential families.

\section{Introduction}

Clustering is a fundamental and key primitive to discover structural groups of homogeneous data, called {\em clusters}, in  data sets.
The most famous clustering technique is the celebrated {\em $k$-means}~\cite{bregmankmeans-2005} that seeks to minimize the sum of intra-cluster variances by prescribing beforehand the number of clusters, $k$.
On one hand, solving the $k$-means problem is {\em NP-hard}~\cite{DasGupta-2008} when the dimension $d>1$ and $k>1$ and various heuristics {\em locally} optimizing the $k$-means objective function like Lloyd's batched $k$-means~\cite{bregmankmeans-2005} have been proposed. 
When $d>1$ and $k>1$, NP-hardness also holds for other  clustering problems like $k$-medoids, $k$-medians and $k$-centers~\cite{kmediankcenter-1984}. 
On the other hand, it is well-known that those center-based clustering problems are fully characterized when $k=1$: For example, the {\em centroid}~\cite{bregmankmeans-2005} is the solution of the $1$-mean, the Fermat-Weber point~\cite{kmediankcenter-1984} the solution of the geometric $1$-median, the circumcenter~\cite{kmediankcenter-1984} the solution of the $1$-center, etc.
Surprisingly, it is  less known that $k$-means can be solved {\em exactly} in 1D by using {\em dynamic programming}~\cite{ClusteringDynamicProg-1973,optkmeans1D-2011} (DP).

In this letter, we first revisit and extend the seminal dynamic programming (DP) paradigm~\cite{ClusteringDynamicProg-1973} for optimally clustering $n$ 1D elements into $k$ pairwise disjoint intervals, the clusters. 
We term clustering with this property: The {\em 1D contiguous} or {\em interval clustering problem}.
We further show how to incorporate constraints on the minimum  and the maximum cluster sizes, and perform model selection (i.e., choosing the appropriate $k$) from the DP table. The generic DP solver requires either $O(n^2kT_1(n))$ time using $O(nk)$ memory or $O(n^2T_1(n))$ time using $O(n^2)$ memory, where $T_1(n)$ is the time requires for solving the corresponding $1$-cluster problem.
Second, we consider two applications that refine the generic DP method: 
In the first application, we report a $O(n^2k)$-time optimal Bregman $k$-means relying on 1D Summed Area Tables~\cite{SAT-1984} (SATs) and also consider the Bregman $\ell_r$-clustering problems~\cite{TotalBregmanClustering-2012}.
In the second application, we consider learning statistical mixture models from independently and identically (iid.) univariate observations by maximizing the complete likelihood:
Using the one-to-one mapping between Bregman divergences and exponential families~\cite{bregmankmeans-2005}, we transform this problem into a series of equivalent 
1D Bregman $k$-means clustering that can be solved optimally by DP for statistical mixtures of {\em singly-parametric exponential families} (like zero-centered Gaussians, Rayleigh or Poisson families). In the general case, we require that the density graphs intersect pairwise  in at most a single point like the Cauchy or Laplacian location families (not belonging to the exponential families) to guarantee optimality.

\section{1D contiguous clustering: Interval clustering\label{sec:1dclustering}}
Let $\bbX$ be a one-dimensional space totally ordered with respect to $<$ (usually, $\bbX=\bbR$), and 
 $\calX=\{x_1, ..., x_n\}\subset \bbX$ a set of $n$ distinct elements.
A clustering of $\calX$ into $k\in\bbN$ clusters partitions $\calX$ into pairwise disjoint subsets $\calC_1\subset\calX, ..., \calC_k\subset\calX$
so that $\calX=\biguplus_{i=1}^k \calC_i$.
Let us preliminary sort $\calX$ in $O(n\log n)$ time, so that we assume $x_1<...<x_n$ in the remainder.

The output of a 1D contiguous clustering is a collection of $k$ intervals $I_i=[x_{l_{i}},x_{r_{i}}]$  (such that $\calC_i= I_i\cap \calX$) that can be 
encoded using $k-1$ {\em delimiters} $l_i$ ($i\in\{2, ...,k\}$) since $r_i=l_{i+1}-1$ ($i<k$ and $r_k=n$) and $l_1=1$: 

\begin{equation}
\underbrace{[x_1 ... x_{l_2-1}]}_{\calC_1}\   \underbrace{[x_{l_2} ... x_{l_3-1}]}_{\calC_2}  ...    \underbrace{[x_{l_{k}} ... x_n]}_{\calC_k}
\end{equation}

To define an optimal clustering among the potential ${{n-1}\choose{k-1}}$ contiguous partitions, we ask to minimize a clustering {\em objective function} or {\em energy function}:

\begin{equation}\label{eq:clusteringcost}
\min_{l_1=1<l_2<...<l_k} e_k(\calX)= \bigoplus_{i=1}^k e_1(\calC_i),
\end{equation}
where $e_1$ denotes the {\em intra-cluster cost} and $\oplus$ is a commutative and associative operator for calculating the {\em inter-cluster cost}.
This framework includes the $k$-means and the $k$-medians ($\bigoplus=\sum$), and the $k$-center~\cite{kmediankcenter-1984} ($\bigoplus=\max$) criteria (and their discrete counterparts: $k$-medoids, etc.) among others.

\subsection{Solving 1D contiguous clustering using DP}
\begin{figure}
\centering
\includegraphics[width=0.8\columnwidth]{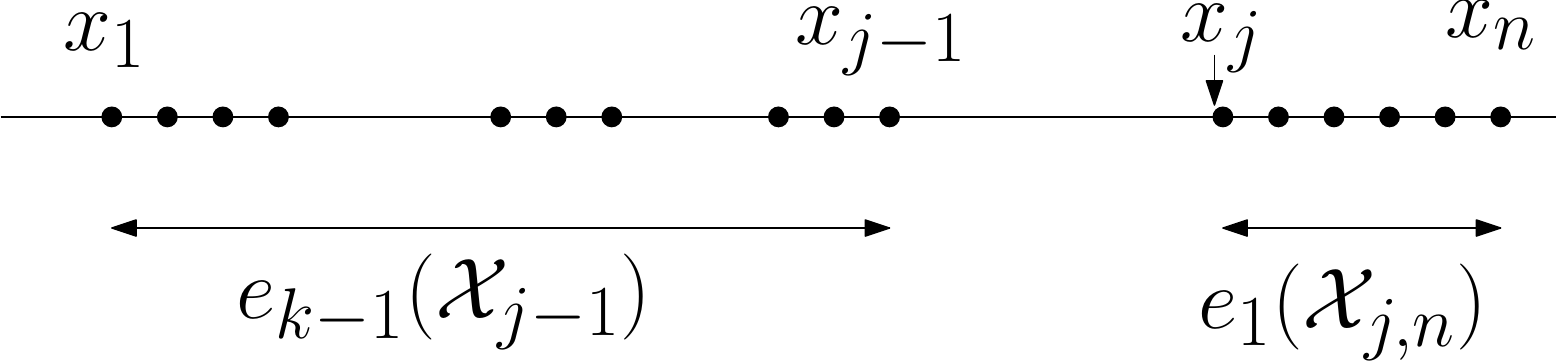}
\caption{The optimal 1D contiguous clustering  is found by dynamic programming by observing that an optimal clustering with $k$ clusters is necessarily found from an optimal  clustering with $(k-1)$ clusters (see text and Eq.~\ref{eq:recDP}).\label{fig:decomp} }
\end{figure}

Recall that after sorting, we have $x_1<...<x_n$.
Let $\calX_{j,i}=\{x_j, ..., x_i\}$ ($j\leq i$) and $\calX_i=\calX_{1,i}=\{x_1, ..., x_i\}$.
We define a $n \times k$ cost matrix $E=[e_{i,j}]$ that stores at entry $(i,m)$ the optimal clustering cost $e_{i,m}=e_{m}(\calX_i)$, where $e_m$ is defined using Eq.~\ref{eq:clusteringcost}.
Similarly, we define a matrix $S=[s_{i,j}]$ of dimension $n \times k$ that stores at position $(i,m)$ the index $j$ of the leftmost point in the $m$-th cluster in ${\mathcal{X}}_i$.
Therefore the global clustering solution shall be found at entry $(n,k)$ with cost $e_{n,k}=e_{k}(\calX)$.

To define the {\em optimality equation} of dynamic programming, we observe that the optimal solution for a 1D contiguous clustering with $m$ clusters can be defined from the solution of an optimal clustering with $(m-1)$ clusters:
Indeed, consider the last cluster interval with left position index $l_m$, say $l_m=j$, as depicted in Figure~\ref{fig:decomp}. 
Then the clustering of the $(m-1)$ first clusters should be an optimal clustering too: namely, the optimal 1D contiguous clustering with $(m-1)$ clusters on subset $\calX_{j-1}$. 
It follows the following recurrence equation:
\begin{equation}\label{eq:recDP}
e_{i,m} = \min_{m\leq j\leq i} \left\{ e_{j-1,m-1}  \oplus e_1(\calX_{j,i})  \right\},
\end{equation}
with  $e_{i,1}=e_1(\calX_i)$  (note that $e_{m,m}=\bigoplus_{l=1}^m e_1(\{x_l\})$ for $1\leq m \leq k$).
We store the argmin of Eq.~\ref{eq:recDP} in matrix $S$ at position $(i,m)$ (entry $s_{i,m}$).
We compute the energy matrix $E$ from left to right columns, and from bottom to top lines.
This yields a $O(n^2 k T_1(n))$-time DP algorithm using $O(n\times k)$ memory, where $T_1(n)$ denotes the time required for computing $e_1(\calX)$: 
Indeed, each of the $n\times k$ entries of $E$ requires $O(n T_1(n))$ time to evaluate Eq.~\ref{eq:recDP}.
 
To recover the optimal clustering, we {\em backtrack} the solution in $O(k)$ time from the $S$ matrix storing the left indexes of the last cluster of the  best solutions:
That is, the left index $l_k$ of the $k$-th cluster is stored at $s_{n,k}$: $l_k=s_{n,k}$.
The cardinality of $\calC_k$ is $n_k=|\calC_k|=n-l_k+1$.
Then we iteratively retrieve the previous left interval indexes at entries $l_{j-1}=s_{l_j-1,j-1}$ for 
$j=k-1, ..., j=1$ with $n_j=|\calC_j|=r_j-l_j+1=l_{j+1}-l_j$ since $r_j=l_{j+1}-1$. Note that $l_j-1=n-\sum_{l=j}^k n_l$ denotes the remaining
number of elements to cluster using $(j-1)$ clusters (thus we also have $l_j-1=\sum_{l=1}^{j-1} n_l$).

Note that when the clustering does not satisfy the 1D contiguous partition property, DP yields {\em anyway} a solution that may not be optimal.
Furthermore, we may consider adding a weight $w_i>0$ to each element $x_i\in\calX$ (and thus assume the $x_i$'s are all distinct).

\subsection{Time versus memory optimization}
By precomputing all the potential intra-cluster costs $e_1(\calX_{j,i})$  in $O(n^2 T_1(n))$ time using an auxiliary matrix $E_1$ of size $n\times n$, we evaluate Eq.~\ref{eq:recDP}  as $e_{i,m} = \min_{m\leq j\leq i} \{ e_{j-1,m-1}  \oplus E_1[j,i]\}$, i.e. in $O(i-m)=O(n)$ time.
Matrix $E_1$ plays the role of a {\em Look Up Table} (LUT), and the time complexity for the DP solver reduces to $O(n^2 k)$ once the LUT matrix $E_1$ has been computed. 

\begin{lemma}
The generic 1D contiguous clustering can be solved optimally using dynamic programming in time $O(n^2 k T_1(n))$ using $O(n\times k)$ memory, or in time $O(n^2T_1(n))$ time using $O(n^2)$ memory.
\end{lemma} 
Note that $T_1=\Omega(n)$ (in fact, usually, $T_1(n)=\Theta(n)$).
In Section~\ref{sec:bregman}, we will further improve the running time to $O(n^2k)$ using $O(nk)$ memory when considering Bregman $k$-means.

\subsection{Adding cluster size constraints}

Let us add constraints on the sizes of clusters.
Let $n_i^-$ and $n_i^+$ denote lower and upper bound constraints on the size of the $i$-th cluster $n_i=|\calC_i|$,
with $\sum_{l=1}^k= n_i^-\leq n$ and $\sum_{l=1}^k= n_i^+\geq n$.
When no constraints are required, we simply add the {\em dummy} constraints  $n_i^-=1$ and $n_i^+=n-k+1$ (all clusters non-empty).
In Eq.~\ref{eq:recDP}, $j$ range from $m$ to $i$.
The $m$-th cluster size $n_m=|\calC_m|=i-j+1$ has to satisfy $n_m^- \leq n_m\leq n_m^+$.
That is, $j\leq i+1-n_m^-$ and $j\geq i+1-n_m^+$.
Clearly, $j$ has also to be greater than $1+\sum_{l=1}^{m-1} n_l^-$ (an optimal solution for the constrained optimal $(m-1)$-clustering).
It follows, that the optimality equation writes as:

\begin{equation}\label{eq:recDP2}
e_{i,m} = \mathop{\min_{\max\{1+\sum_{l=1}^{m-1} n_l^-,i+1-n_m^+\}\leq j}}_{j\leq i+1-n_m^-} \left\{ e_{j-1,m-1}  \oplus e_1(\calX_{j,i})  \right\},
\end{equation}

For example, a balanced clustering may be obtained by setting $n_i^-=\floor{\frac{n}{\lambda k}}$ and $n_i^+=\ceil{\frac{\lambda n}{k}}$ for some
 $\lambda\in\bbN$.

\subsection{Choosing the appropriate $k$: Model selection}
\begin{figure}
\centering
\includegraphics[width=0.8\columnwidth]{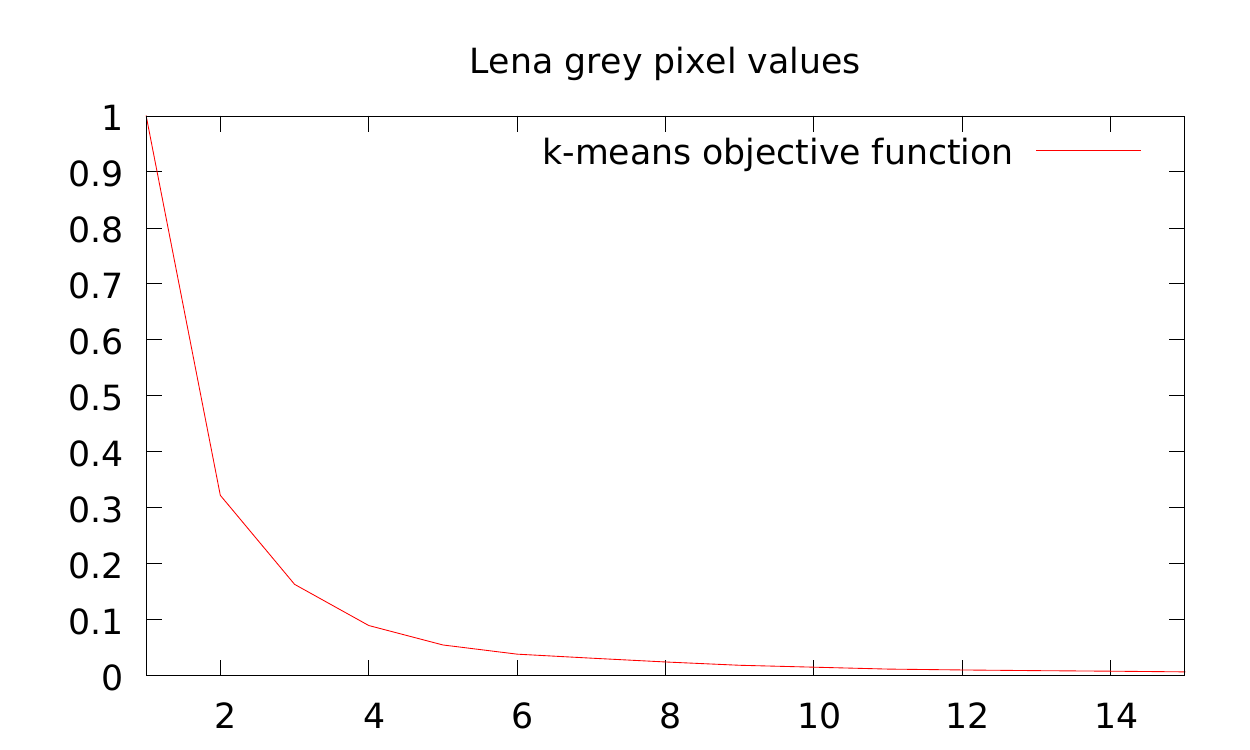}
\caption{Plot of function $m(k)=e_k(\calX)/e_1(\calX)$ for the optimal $k$-means for $k\in[1,15]$.
\label{fig:modelselection}}
\end{figure}

The task of clustering data set $\calX$ asks also to find the appropriate number of clusters~\cite{xmeans-2000}: $k$.
Clearly, the more clusters we allow and the less costly the objective function $e_k(\calX)$ is, but the more complex the clustering model to encode.
Observe that function $m(k)=\frac{e_k(\calX)}{e_1(\calX)}$ is {\em monotonically decreasing} with $k$ and reaches a minimum when $k=n$ (e.g., $0$ for the Euclidean $k$-means) as depicted in Figure~\ref{fig:modelselection} (see \ref{sec:mixture} for an explanation of the data-set).
Thus we have to perform some kind of {\em model selection}~\cite{xmeans-2000} by choosing the {\em best model} among all potential models (with number of clusters ranging from  $1$ to $n$).
The canonical {\em regularized objective clustering cost}~\cite{xmeans-2000} is $e_k'(\calX)=e_k(\calX)+f(k)$ where $f(k)$ is the cost function of choosing a model with $k$ clusters.
We can compute the best model minimizing $e_k'(\calX)$ by computing for the DP table entries for the {\em last matrix row} of $E$ (indexed by $n$, with columns $k$ ranging from $1$ to $n$) the regularized cost. 
To compute the last row, we iteratively solve DP for $k=n, n-1, ..., 1$ and avoid redundant computations by checking whether entry $E[i,j]$ has already been computed or not.
We then choose $k=\argmin_{g\in\{1, ..., k\}} e_g'(\calX)$ by scanning the last row with column ranging from $k=1$ to $k=n$.

\subsection{A Voronoi condition for optimal center-based clustering}
Center-based clustering methods like $k$-means, $k$-medians or $k$-centers store for each cluster $\calC_j$ a {\em prototype} $p_j$, the cluster center.
For discrete center-based clustering, the prototypes $p_j$'s are required to belong to the respective $\calC_j$'s.
The $\ell_r$ {\em center-based clustering objective function} asks to minimize:
\begin{equation}
\sum_{i=1}^n  w_i \min_{j=1}^k d^r(x_i,p_j)=\sum_{j=1}^k \sum_{x_l\in\calC_j} w_l d^r(x_l,p_j),
\end{equation}
where $d(\cdot,\cdot)$ is a {\em dissimilarity measure function} ({\em not} necessarily a distance).
We do not take the $\frac{1}{r}$ power of the sum since it changes the value of $e_1$ but not the argmin (prototype).
Note that in 1D, $\ell_s$-norm distance is always $d_s(p,q)=|p-q|$, independent of $s\geq 1$.
Thus the intra-cluster cost $e_1(\calC_j)$ of a  $\ell_r$ center-based clustering has to solve  the following minimization problem:
$e_1(\calC_j)=\min_{p_j} \sum_{x_l\in\calC_j} w_l d^r(x_l,p_j)$ and retrieve the $j$-th cluster prototype by $p_j=\argmin_{p_j} \sum_{x_l\in\calC_j} w_l d^r(x_l,p_j)$.

In order for DP to return the optimal clustering, we need to assume that we have the 1D contiguous clustering property.
For Euclidean $k$-means, this was proved in~\cite{kmeans1D-1958}. 
In general, consider the {\em Voronoi cell} of prototype $p_j$ of $\calC_j$:
\begin{equation}
V(p_j)=\{x\in\bbX\ :\ d^r(x,p_j) \leq d^r(x,p_l)\ \forall l\in\{1, ...,k\}  \}.
\end{equation}
Since $x^r$ is a monotonically increasing function on $\bbR^+$, it is equivalent to 
$V'(p_j)=\{ x\in\bbX\ : \  d(x:p_j)  <    d(x:p_l)\}$.
A sufficient condition is to prove that for {\em all} potential choices of the $k$ cluster prototypes $\calP=\{p_1, ..., p_k\}$ the induced 1D dissimilarity Voronoi diagram is made of {\em connected Voronoi cells}. A $2$-clustering displays the Voronoi bisector.
We now consider two case studies to illustrate and refine the DP method.

\section{Optimal 1D Bregman clustering\label{sec:bregman}}

The $\ell_r$-norm Bregman center~\cite{TotalBregmanClustering-2012} is defined for $d(p,q)=B_F(p:q)$, where $B_F(p:q)$ is a univariate Bregman divergence~\cite{bregmankmeans-2005}:
\begin{equation}
B_F(p:q)=F(p)-F(q)+(p-q)F'(q),
\end{equation}
induced by a strictly convex and differentiable function $F$. When $F(x)=x^2$, we recover the squared Euclidean distance.
Bregman divergences are {\em not} metric~\cite{BVD-2010}, since they violate the triangular inequality and are {\em asymmetric} except when $F(x)=\lambda x^2$ for $\lambda>0$.

For Bregman $k$-means, the {\em Bregman information}~\cite{bregmankmeans-2005} of a cluster generalizes the notion of cluster variance.
It is the {\em intra-cluster sum of  Bregman divergences} (Bregman $k$-means, for $r=1$):

\begin{equation}
e_1(\calC_j)=  \min_{p_j} \sum_{x_l\in\calC_j} w_l B_F(x_l : p_j).
\end{equation}
The cluster prototype~\cite{bregmankmeans-2005} is $p_j=\frac{1}{\sum_{x_l\in\calC_j} w_l}\sum_{x_l\in\calC_j} w_lx_l$ and the Bregman information is~\cite{SymmetrizedBDCentroid-2009}: 
$e_1(\calC_j) = \left(\sum_{x_l\in\calC_j} w_l\right) (p_j F'(p_j)-F(p_j)) +\left(\sum_{x_l\in\calC_j} w_l F(x_l)\right) - F'(p_j)\left(\sum_{x\in\calC_j} w_l x\right)$. 
 Observe that the Bregman information   relies on three sums
 $\sum_{x_l\in\calC_j} w_l$, $\sum_{x\in\calC_j} w_l x$ and $\sum_{x_l\in\calC_j} w_l F(x_l)$ that can be preprocessed using {\em Summed Area Tables}~\cite{SAT-1984} (SATs) since $\calC_j$ is a contiguous cluster.
That is, by computing all the {\em cumulative sums} $S_1(j)=\sum_{l=1}^j w_l$, $S_2(j)=\sum_{l=1}^j  w_lx_l$, and $S_3(j)=\sum_{l=1}^j  w_lF(x_l)$  in $O(n)$ time at preprocessing stage,
we can evaluate the Bregman information $e_1(\calX_{j,i})$ in constant time $O(1)$.
For example, $\sum_{l=j}^i w_l F(x_l)=S_3(i)-S_3(j-1)$ with the convention that $S_3(0)=0$.

The Voronoi cells of prototypes are defined by $V'(p_j)=\{ x\in\bbX\ : \  B_F(x:p_j) < B_F(x:p_l)\}$.
Since {\em Bregman Voronoi diagrams} have connected cells~\cite{BVD-2010}, it follows that the 1D hard $\ell_r$ Bregman clustering satisfies the contiguous interval property, and therefore DP yields the optimal solution. 
A similar argument directly hold for the Bregman $k$-center that is also the limit case of $\ell_r$ Bregman clustering when $p\rightarrow\infty$.

\begin{lemma}
The  1D $\ell_r$ Bregman clustering and Bregman $k$-center can be solved exactly using dynamic programming in $O(n^2 k T_1(n))$ time using $O(n\times k)$ memory, where $T_1(n)$ denotes the time to solve the case $k=1$ for $n$ elements.
The optimal Bregman $k$-means can be solved in $O(n^2k)$ time.
\end{lemma}

\section{Mixture learning by hard clustering\label{sec:mixture}}
Statistical mixtures are semi-parametric probability models often met in practice.
Consider a finite {\em statistical mixture} $M$ with $k\in\bbN$ components.  
The probability measure $m$ of $M$  with respect to a dominating measure $\nu$ (usually the Lebesgue or counting measure) can be written as:
\begin{equation}
m(x;\Omega) = \sum_{i=1}^k \alpha_i p(x;\Theta_i), x\in\bbX,
\end{equation}
with $\alpha=(\alpha_1, ..., \alpha_k)\in\Delta_{k-1}$ a normalized positive weight vector belonging to the $(k-1)$-dimensional {\em probability simplex}, 
$\Theta=(\Theta_1, ..., \Theta_k)$, $\Omega=(\alpha,\Theta)$ and $\bbX$ the support of the distribution.
Let $D=\dim(\Theta_i)\in\bbN$ denote the number of scalar parameters indexing the probability family $\calF=\{p(x;\Theta) \st \Theta \in \bTheta  \}$, called the  {\em order}.
Mixture $m$ is defined by a vector $\Omega\in\bOmega\subseteq\bbR^g$ with $g=k(D+1)-1$, and $\bTheta$ is called the {\em parameter space}.
Mixtures are inferred from data usually using the Expectation-Maximization algorithm~\cite{bregmankmeans-2005}.
Since EM locally maximizes the {\em incomplete likelihood}~\cite{bregmankmeans-2005} and is often trapped into a local maximum, we need some proper mixture parameter initialization or several guided restarts to hopefully reach the optimal solution. 
On the other hand, maximizing the {\em complete log-likelihood} $l_c$ for a iid. observation data-set $\calX$ amounts to maximize~\cite{kmle-2012}:
\begin{equation}
l_c(\calX; L,\Omega) = \sum_{i=1}^n  \log(\alpha_{l_i} p(x_i;\theta_{l_i})),
\end{equation}
where  $L=\{l_i\}_i$ denotes the hidden labels of the $x_i$'s.
Thus maximizing the complete likelihood is equivalent to minimizing the following objective function:
\begin{equation}\label{eq:clopt}
\max l_c \equiv \min_{\theta_1,...,\theta_k} \sum_{i=1}^n  \min_{j=1}^k  (-\log p(x_i;\theta_{j}) -\log \alpha_{j}).
\end{equation}
This is a hard clustering problem for the dissimilarity function $d(x,(\alpha,\theta))=-\log p(x;\theta) -\log \alpha$ (given fixed $\alpha$).
As proved in~\cite{kmle-2012}, the cluster weights $\alpha_j$'s are then updated as the cluster proportion of observations, and the algorithm reiterates by solving Eq.~\ref{eq:clopt}. Initially, we choose $\alpha=\frac{1}{k}(1, ..., 1)$.

Let the {\em additively-weighted minus log-likelihood Voronoi cell} be defined by 
$V(p_j)=\{x\in\bbX \st -\log p(x;\theta_j) -\log \alpha_j  \leq -\log p(x;\theta_l) -\log \alpha_l\}$.
In order for DP to return the optimal solution, we need to assert the contiguity property.
Using the one-to-one mapping between exponential families~\cite{BrownExpFam-1986,ef-flashcards-2009} and Bregman divergences~\cite{bregmankmeans-2005}, it turns out that the optimization problem of Eq.~\ref{eq:clopt} yields an equivalent additively-weighted Bregman  $k$-means problem (and additively-weighted Bregman Voronoi cells are connected~\cite{BVD-2010}).
Thus when the order of the exponential family is $D=1$, we have the contiguity property and DP returns the optimal solution.
This works also for curved exponential families with one free parameter like the family of Gaussian distributions $\calF=\{N(\mu,\mu^2)\ : \ \mu\in\bbR\}$.
In general, the contiguity property holds when density graphs in $\calF$ are pairwise intersecting at exactly one point of the support $\bbX$.
For example, some (unimodal) {\em location families} with density $\calF=\{f(x;\mu)=\frac{1}{\sigma}f_0(\frac{x-\mu}{\sigma}), \mu\in\bbR\}$ for a prescribed value of $\sigma>0$ and a standard density $f_0(x)$ (e.g., isotropic gaussian densities $N(\mu_1,\sigma)$ and $N(\mu_2,\sigma)$ intersect at $x=\frac{\mu_1+\mu_2}{2}$).
This includes location Cauchy distributions  and location Laplacian distributions (both not belonging to the  exponential families~\cite{BrownExpFam-1986}) among others.
Note that $1$-order exponential families may have pairwise densities intersecting in more than one point (like the family $\calF=\{N(0,\sigma), \sigma\in\bbR^+\}$) but after reparameterization by their sufficient statistic~\cite{BrownExpFam-1986} $y_i=t(x_i)$, data-set $\calY=\{y_i\}_i$ satisfies the contiguous property.

\begin{figure}
\centering
\includegraphics[width=0.8\columnwidth]{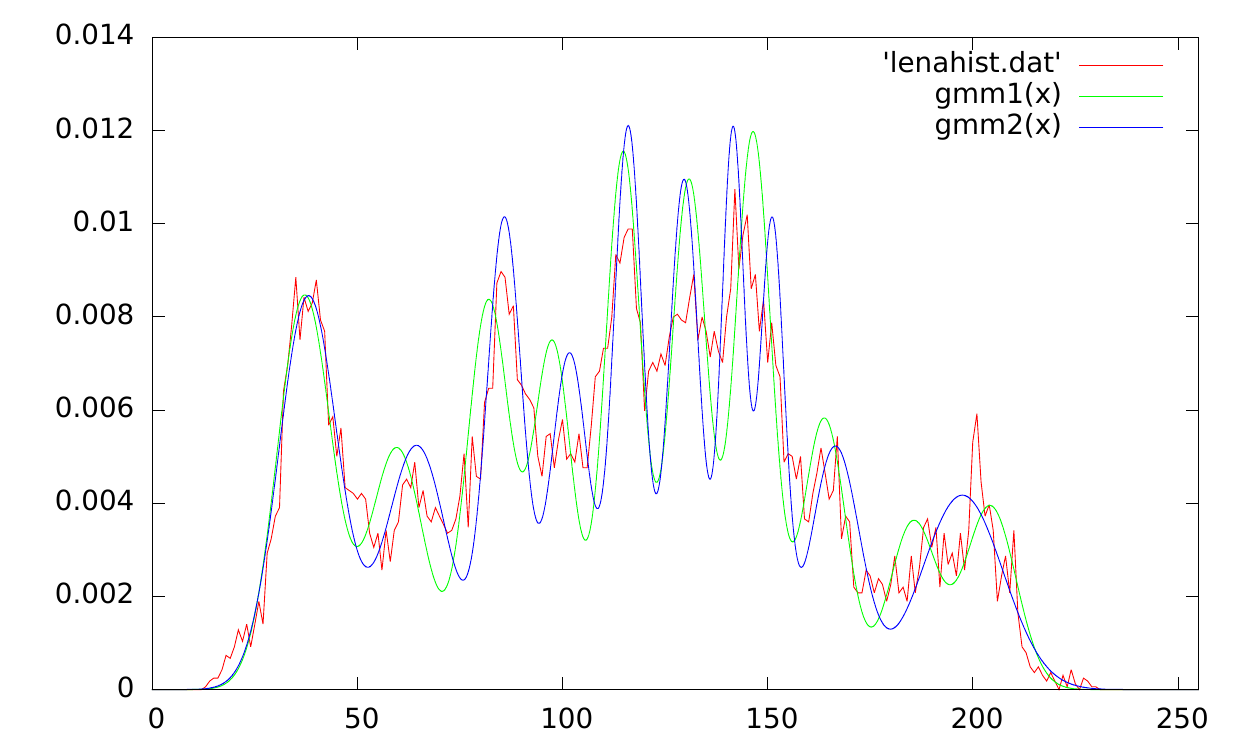}
\caption{1D GMMs with $k=10$ components maximizing the complete data likelihood of the intensity histogram of {\tt lena} image: $\mathrm{gmm}_1$ retrieved from an optimal Euclidean $k$-means, and $\mathrm{gmm}_2$ allowing different standard deviations. 
The average complete data log-likelihood of $\gmm_1$ is $-3.075$ and that of $\gmm_2$ is $-3.039$ (better than the one for $\gmm_1$). 
\label{fig:res}}
\end{figure}

Consider fitting a Gaussian Mixture Model (GMM) on the intensity histogram of the renown {\tt lena} color image.
For each pixel, we compute its grey value and add a small perturbation noise to ensure that we get distinct $x_i$'s
(alternatively, without adding noise, we set the weight $w_i$ of $x_i$ as the proportion of pixels having grey value $x_i$). 
We then compute the optimal Euclidean 1D $k$-means for $k=10$ (it corresponds to fitting a 1D GMM  $\gmm_1$ with Gaussian components having identical\footnote{Once we get the optimal Euclidean cluster decomposition, we fit in each cluster its maximum likelihood estimator (MLE) mean and standard deviation from the cluster data, and set $\alpha$ as the relative proportion of points.} standard deviation), and calculate
the 1D GMM $\gmm_2$ allowing different standard deviations. 
In that case, we do {\em not} have the contiguous clustering property (densities pairwise intersect in two points)  and DP may {\em not} yield the optimal clustering (give prescribed weights).
However, in this case, we experimentally obtained a better GMM.
The results are illustrated in Figure~\ref{fig:res}.
For model selection in mixtures, to choose the optimal $k$, we  use the {\em Akaike Information Criterion}~\cite{AIC-1997} (AIC):
$\AIC(x_1, ..., x_n) =-2 l(x_1, ..., x_n) + 2k + \frac{2k(k+1)}{n-k-1}$.
Other criteria like the Bayesian Information Criterion (BIC), Minimum Description Length (MDL), etc can also be used.

\section{Conclusion}
We first described a clustering algorithm based on dynamic programming  (whose seminal idea was briefly outlined in Bellman's $2$-page paper~\cite{ClusteringDynamicProg-1973} in 1973) that computes  the generic optimal 1D contiguous clustering  either in $O(n^2 k T_1(n))$-time using $O(nk)$ memory, or in $O(n^2T_1(n))$ time using $O(n^2)$ memory, where $T_1(n)$ denotes the time required for solving the case $k=1$ on $n$ scalar elements. We then extended the method to incorporate cluster size constraints and show how to perform model selection from the DP table. This algorithm solves optimally and generically 1D $k$-means, $k$-median and $k$-center among others. 
Second, we reported two tailored center-based clustering applications of the optimal 1D contiguous clustering: 
(1) Bregman $k$-means and $k$-centers clustering, and (2) learning statistical mixtures maximizing the complete likelihood provided that 
(a) their densities belong to  a $1$-order exponential family or (b) their density graphs pairwise intersect in one point.
For Bregman $k$-means, we showed how to use Summed Area Tables (SATs) to further speed the DP solver in $O(n^2k)$-time using $O(nk)$ memory.


\end{document}